\documentclass[newstyle]{rmaa}
\usepackage{rmaacite}

\def\LCDM{$\Lambda$CDM}
\def\msun{${\rm M}_{\odot}$}
\def\cnfw{$c_{NFW}$}
\newcommand{\kmsmpc}{\, \rm{km}\,  \rm{s}^{-1}\, \rm{Mpc}^{-1}}
\def\apj{ApJ}

\def\apjl{ApJL}
\def\apjs{ApJS}

\def\mnras{MNRAS}

\title{Evaluating the disk formation paradigm in the $\Lambda$CDM framework:
constraints from the Milky Way}
\author{Rachel S. Somerville
\affil{Department of Astronomy, University of Michigan} }

\shortauthor{R.S. Somerville}
\shorttitle{Evaluating the disk formation paradigm}

\suppressfulladdresses

\abstract{We investigate the properties of a galaxy similar to the
Milky Way within the context of standard disk formation theory in a
\LCDM\ universe. Using the standard assumption that baryons conserve
specific angular momentum when they collapse, we conclude that the
\emph{mean} properties of the model galaxies are in good agreement
with the Milky Way and other similar spiral galaxies, but the
predicted scatter in disk scale lengths may be too large. A model in
which half of the initial specific angular momentum is transfered to
the dark matter may produce a smaller scatter, if very compact disks
are unstable and evolve into spheroids or early type galaxies.}

\resumen{}


\begin{document}
\maketitle

\section{Introduction}
Cold Dark Matter (CDM) seems to provide a very successful paradigm for
explaining many different kinds of observations on large scales, but
suffers from several problems on small scales. Perhaps the most
worrisome of these is the ``cusp'' problem: it seems that the radial
profiles of dark matter halos produced in cosmological simulations
based on CDM are inconsistent with the observed rotation curves of at
least some dwarf and low surface brightness galaxies (e.g. van den
Bosch \& Swaters 2001 and references therein). It is important to
establish whether or not this problem is peculiar to this particular
class of galaxies. It is more difficult to assess whether the rotation
curves of luminous, high surface brightness galaxies are consistent
with CDM dark matter halos, because in these galaxies, baryons
contribute significantly to the gravitational force in the central
part of the galaxy, where rotation curves are observed. It is
therefore expected that the inner dark matter profile is significantly
modified by the collapse of the baryons. It is possible to calculate
the effect of this baryon-induced ``contraction'' on the dark matter
halo using a well-established analytic formalism. However, the large
degeneracies due to the many unknown parameters make it difficult to
obtain strong constraints from the observed rotation curves of most
luminous galaxies.

The Milky Way galaxy offers a special opportunity to investigate this
question. We know about the dynamical properties of our Galaxy in much
greater detail and over a larger range of scales than any other
galaxy. For example, the mass profile of our Galaxy as a function of
radius is constrained by velocity measurements from scales of a few pc
(from stellar velocities) to 100 kpc (from satellite galaxies). Also,
observations of microlensing events towards the Galactic bulge place
strong lower limits on the mass of \emph{baryonic} material within
about 3 kpc. In Klypin, Zhao \& Somerville (2002; KZS02; see also the
contribution by A. Klypin in this volume), we showed that these
combined data place very strong constraints on the parameters of the
Milky Way Galaxy and its dark matter halo. KZS02 concluded that,
within the framework of the popular \LCDM\ cosmological model: 1) the
Milky Way must occupy a halo with a total mass in the range 1--$2
\times 10^{12}$ \msun\ 2) half of the baryons within the virial radius
of this halo must have been ejected 3) standard disk formation models,
in which the gas conserves its specific angular momentum during
collapse, have difficulty obeying the combined microlensing and
dynamical constraints. However, if angular momentum is
\emph{transfered} from the baryons to the dark matter, the dark matter
gains angular momentum and so moves outward. The inner dark matter
``cusp'' is flattened out, leaving more room for baryons in the inner
part of the Galaxy. KZS02 concluded that a model in which about half
of the initial specific angular momentum was lost by the baryons could
accommodate all of the data.

This brings up several further questions. How typical is our Galaxy?
Does it lie near the mean of the distribution of objects predicted by
the theory, or is it an outlier? Is it reasonable that such a large
fraction of the baryons could be ejected from a relatively massive
halo?  Are the photometric properties of a ``Milky Way'' produced in
this framework consistent with observations? Is the distribution of
stellar ages and metallicities in such a model consistent with
observations in the Galaxy? These questions will be addressed in a
companion paper to KZS02 (Somerville, Klypin \& Zhao, in prep). Here,
we will briefly address a few of these questions.

\begin{figure}
\begin{center}
\end{center}
\includegraphics[width=\columnwidth]{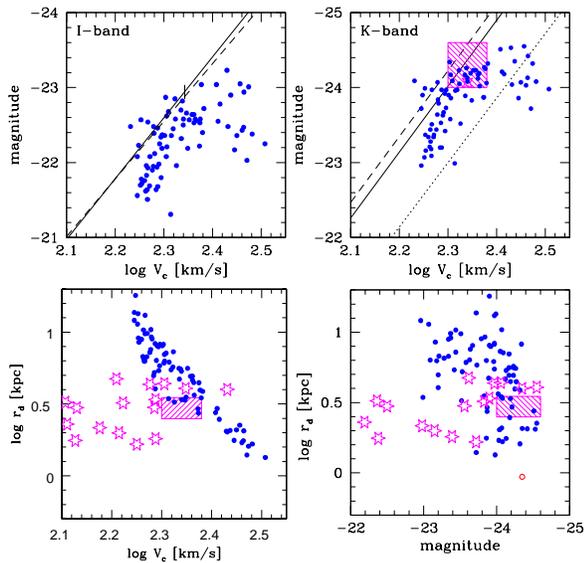}
\caption{\footnotesize Fundamental plane relations predicted by semi-analytic
models, for an ensemble of ``Milky Way'' mass galaxies (total mass of
baryons plus dark matter $10^{12} {\rm M}_{\odot}$), assuming that the
specific angular momentum of the baryons is conserved. Large star
symbols show the observed relations for the Ursa Major sample of
normal spirals from Verheijen (2001). Shaded areas indicate the
acceptable range of values for the Milky Way.  Dots show the model
galaxies, with filled symbols indicating galaxies that should be
globally stable, and open symbols showing galaxies that may be
unstable to bar/bulge formation.
\label{fig:notrans}}
\end{figure}

\begin{figure}
\begin{center}
\end{center}
\includegraphics[width=\columnwidth]{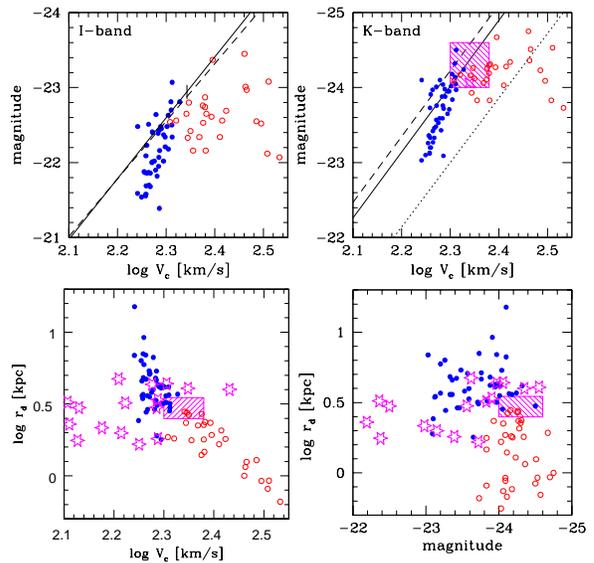}
\caption{\footnotesize The same as Fig.~\protect\ref{fig:notrans}, except that
angular momentum transfer is included using the formalism developed by
KZS02. Note that while the agreement with the observational locus is
better, a large number of unstable disks are predicted (shown by open
symbols).
\label{fig:trans}}
\end{figure}

\section{Properties of Model ``Milky Ways''}
Our chosen tool for this investigation is a semi-analytic model of
galaxy formation.  These models treat the hierarchical history of
galaxy formation using a ``merger tree'', and include recipes for gas
cooling, star formation, and supernova feedback.  We use an updated
version of the models presented in Somerville \& Primack (1999) and
Somerville, Primack \& Faber (2001); see those references for details.
New aspects of the model used here include 1) realistic dark matter
halo profiles 2) more detailed modelling of disk formation, including
the contraction of the halo, based on the ``adiabatic invariant''
formalism (Blumenthal et al. 1986; Flores et al. 1993; Mo, Mao \&
White 1998). Dark matter halos are assumed to follow the universal
Navarro-Frenk-White (NFW) profile (Navarro, Frenk \& White 1997) and
are characterized by the NFW ``concentration'' parameter \cnfw. The
angular momentum of a dark matter halo is characterized by the
dimensionless spin parameter $\lambda$, and spin parameters are chosen
randomly from a log-normal distribution (Bullock et al. 2001a). The
adiabatic invariant formalism then allows us to calculate, at a given
redshift, the exponential disk scale length $r_d$ and the maximum
rotation velocity $V_{\rm max}$ as a function of the halo parameters
\cnfw\ and $\lambda$, and the fraction of baryons that ends up in the
disk+bulge of the galaxy, $f_{\rm gal}= (m_{\rm disk}+m_{\rm
bulge})/(f_b M_{\rm vir})$, where $m_{\rm disk}$ and $m_{\rm bulge}$
are the mass of the disk and bulge, $f_b$ is the universal baryon
fraction, and $M_{\rm vir}$ is the virial mass of the halo. In the
semi-analytic models, $f_{\rm gal}$ is determined by the efficiency of
cooling, star formation, and gas ejection by feedback, and varies from
halo to halo depending on its formation history.

We simulate a large ensemble of halos with a mass of $10^{12}$ \msun,
as in the fiducial Milky Way model of KZS02. The assumed cosmology is
a ``standard'' \LCDM\ model with $\Omega_m=0.3$,
$\Omega_{\Lambda}=0.7$, $\sigma_8=1$, and $H_0 = 70 \kmsmpc$. The
three main free parameters in our model are the efficiency of star
formation and supernova feedback, and the effective yield of heavy
elements. These parameters are adjusted by requiring the average mass
of the disk+bulge of the ``Milky Way'' to be close to $5 \times
10^{10}$ \msun, as favored by the dynamical arguments, the gas
fraction to be close to ten percent, and the mean stellar metallicity
to be close to solar, as observed. These conditions are achieved
easily, with physically reasonable values of the parameters.

The ``Fundamental Plane'' scaling relations obtained for this ensemble
of Milky Way mass galaxies are shown in Fig.~\ref{fig:notrans}, for
the standard assumption of conservation of specific angular
momentum. Here, we have assumed a one-to-one relationship between halo
mass and \cnfw\ (using the model of Bullock et al. 2001b), so the
scatter comes from the range of values of $f_{\rm gal}$, the
distribution of spin parameter $\lambda$, and the spread in
mass-to-light ratio caused by the variation in star formation and
enrichment history. Note that although much of the scatter at fixed
halo mass moves galaxies \emph{parallel} to the Tully-Fisher relation,
there are outliers at high circular velocity. The mean properties of
the ``Milky Way'' galaxies in the ensemble are in good agreement with
the observational constraints. However, in comparison with a larger
sample of spiral galaxies, it seems the predicted scatter in size at
fixed $V_{\rm max}$ or magnitude may be too large.

Predictions including angular momentum transfer of a factor of
$\sim2$, as proposed by KZS02, and worked out using the formalism
presented there, are shown in Fig.~\ref{fig:trans}. We see that
including this effect produces galaxies with smaller scale radii but
with nearly the same maximum rotation velocity. Many extremely compact
galaxies are now produced. However, these objects are very unlikely to
be stable. We have indicated with open symbols the objects that are
expected to be unstable to formation of a bar and/or bulge, according
to the condition $\varepsilon_m \equiv [V_{\rm max}/(GM_d/r_d)]^{1/2}
< 0.75$ (see Mo et al. 1998). If these objects are removed, the mean
of the distribution is still in good agreement with the observations,
while the width of the distribution is narrower, in better agreement
with the global observed distributions. It remains highly uncertain,
however, how accurate this simple stability condition really is, what
the appropriate threshold value should be, and what happens to
unstable objects.
  
\section{What about the scatter in concentration at fixed mass?}
It is well-known that there is actually a significant scatter in the
halo concentration at fixed mass in cosmological simulations
(e.g. Bullock et al. 2001b). Most investigations of disk properties
have ignored this, as we have done above. Recently, it has been
demonstrated that this scatter comes from variation in the
\emph{formation history} of the halos, with early-forming halos having
high values of \cnfw\ and late-forming halos having smaller values
(Wechsler et al. 2002). When we include this correlation in our
semi-analytic merger trees using the scaling found by Wechsler et
al. (2002), we find noticeable correlations between the halo
concentration and many observable properties of the galaxies, such as
color, gas fraction, and stellar age and metallicity (see the
contribution by Wechsler in this volume). Similarly, the expected
connection between halo mass accretion history and rotation curve
shape was already pointed out some time ago by Firmani \& Avila-Reese
(2000). These results suggest that the properties of galactic disks
are not determined only by their spin parameters, as has often been
emphasized, but that halo concentration is also an important factor.

%\clearpage

\end{document}